# Strengthening of Indian Ocean teleconnections permits predictions of springtime rainfall in SE Australia

Stjepan Marčelja

Research School of Physics, The Australian National University, Canberra, Australia

Email: stjepan.marcelja@anu.edu.au
Postal address: 34 McGivern Cres., Canberra, ACT 2902

**Abstract.** Rainy years and dry years in SE Australia are known to be correlated with sea surface temperatures in the specific areas of the Indian Ocean. While over the past 100 years the correlation had been both positive and negative, it significantly increased in strength since the beginning of the 21st century. Over this period, Indian Ocean sea surface temperatures during the winter months, together with the El Niño variations, contain sufficient information to accurately hindcasts springtime rainfall in SE Australia. Using deep learning neural networks trained on the early 21st century data we performed both hindcasting and simulated forecasting of the recent spring rainfall in SE Australia, Victoria and South Australia. The method is particularly suitable for quantitative testing of the importance of different ocean regions in improving the predictability of rainfall modelling. The network hindcasting results with longer lead times are improved when current winter El Niño temperatures in the input are replaced by forecast temperatures from the best dynamical models. High skill is limited to the rainfall forecasts for the September/October period constructed from the end of June.

## 1 Introduction

For more than 50 years Indian Ocean sea surface temperatures (SST) have been known to affect the rainfall over the Australian continent. A comprehensive history of the earlier research is available in Ummenhofer et al. (2008). The present investigation is an attempt to extend the earlier analyses of the influence of Indian Ocean teleconnections affecting the rainfall in SE Australia (Cai and Cowan 2008, Ummenhofer et al. 2009, Cai *et al.* 2011) to the level of practically useful quantitative predictions.

When listing the most important measures of the ocean conditions, the Indian Ocean Dipole (IOD) is accepted as one of the major climate drivers over the large regions of the three continents at its boundaries. IOD index is defined as the difference between sea surface temperature anomaly at the western side (50$^o$E to 70$^o$E and 10$^o$N to 10$^o$S) and the eastern side (90$^o$E to 110$^o$E and 0$^o$S to 10°S) of the basin. When in its negative phase, westerly winds at tropical latitudes accumulate warmer surface water north of Australia leading to increased rainfall.

Another measure of the SST of the Indian Ocean to the north-west of Australia has an even stronger influence on the rainfall in the south-east of the continent. The meridional temperature gradient, defined in Ummenhofer et al. (2009) as the difference of SST anomalies between the regions *sI* (centred at 30°S, 95°E) and *eI* (centred at 10°S, 110°E) strongly correlates with the rainfall during both dry and wet years. In this study we use HadSST 4.0 data to obtain the values for IOD and the *sI* and *eI* ocean regions. Closely linked to Indian Ocean conditions, subtropical ridge position and intensity are also strongly correlated with the rainfall in SE Australia (Cai *et al*. 2011a, 2011b, Timbal and Drosdowsky 2012).

Pacific Ocean conditions, as reflected in the phase of the El Niño-Southern Oscillation, influence the weather throughout Australia. We used the monthly Niño34 anomaly data based on NOAA ERSST V5 from NOAA Climate Prediction Center. Finally, monthly rainfall data for SE Australia, South Australia and Victoria are kindly provided by the BOM.

The method used consists of two steps. In the first step, we search for climate variables with a significant correlation with the rainfall in SE Australia. In this step we explore the correlation between SST at the known indicative regions of the Indian Ocean and the rainfall over the last 70-year period and different annual seasons.

In the second step, the promising data streams are used as an input into a machine learning artificial neural network. The method is well-known, and in climate research has been used to predict both El Niño (e. g. Nooteboom et al. 2018) and Indian Ocean Dipole (e. g. Ratnam et al. 2020) changes. We train the network with a part of the rainfall data and stop when it achieves the best match with the validation part of the data. This empirical procedure eventually selects the best input variables and the limited window of accurate predictions.

## 2 Changes in teleconnections between the Indian ocean and rainfall in SE Australia

The variables that showed strongest correlation with the rainfall in SE Australia were the temperature anomalies in the *eI*, *sI* and East IOD regions and the Niño34 anomaly. Southern Annular Mode was also explored but did not help in the medium-term hindcasting tests, most likely because of too rapid changes.

As described in earlier studies (e. g. Cai et al 2011b, Timbal and Drosdowsky 2012) correlations exhibit large variations with annual seasons. They are particularly strong in springtime, when Indian Ocean conditions influence subtropical ridge intensity, which in turn influences rainfall. Limited exploration of the geopotential pressure time series did not improve the accuracy of the hindcasts, possibly because the information is already contained in the Indian Ocean conditions. Nevertheless, optimal inclusion of the subtropical pressure ridge position and intensity data in the input would likely improve the forecasting skill.

Important variation of the correlations between the major climate drivers on the decadal timescale was noted by many authors. During the cold phase of the Inter-decadal Pacific Oscillation, spring rainfall in SE Australia is more strongly related to IOD and ENSO and hence more predictable (Lim *et al.* 2017a).

As an example, decadal changes in the correlation between August Meridional gradient and September/October rainfall in SE Australia are shown in Fig. 1. There are periods of both positive and negative correlation, ending with a high correlation period in the 21st century. The variation of the correlation scales with the Pacific Decadal Oscillation index, particularly during the last 50 years (Fig. 1). It does not appear to be related to decadal changes in the Indian Ocean (Han et al. 2014). The 21st century correlation increase is concentrated in winter and spring seasons as illustrated in Fig. 2.

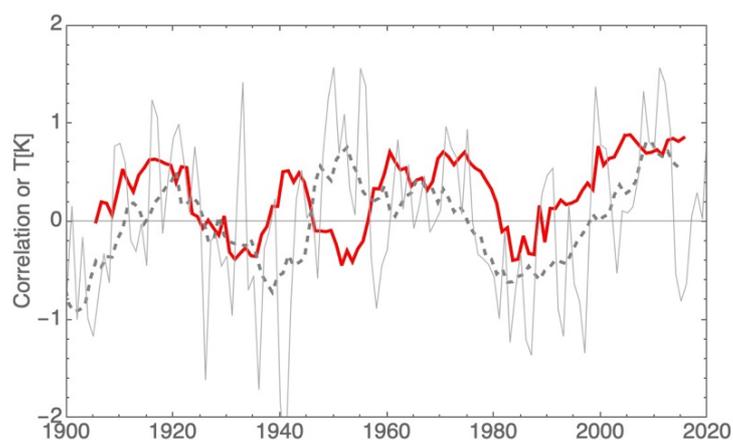

Fig. 1.
Correlation between the Meridional gradient in the Indian Ocean during August and the rainfall in SE Australia in September and October, as evaluated with a 10-year sliding window (red). Also shown is the (negative) Pacific Decadal Oscillation index, together with the 10-year sliding average (grey). The similarity over the whole range is modest, but over the last 50 years becomes very strong (correlation 0.84).

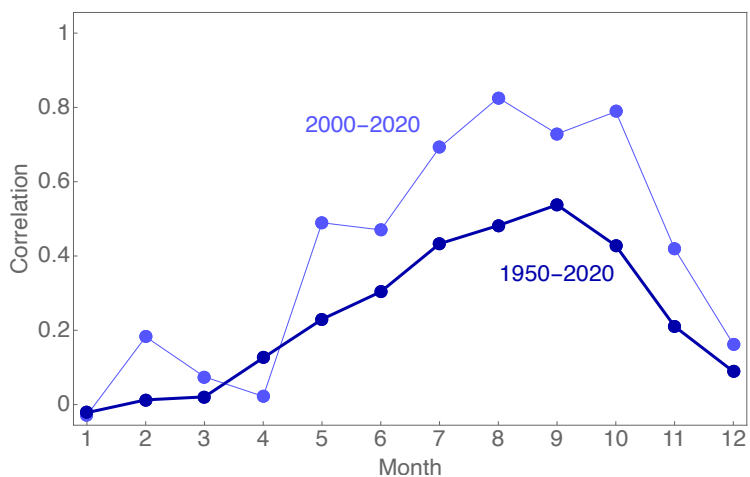

Fig. 2.
An example of the increasing correlation between Indian Ocean SST and the rainfall in SE Australia. The figure shows the correlation between the Meridional gradient and the rainfall in SE Australia during the same month. Correlation is stronger over the more recent period.

In the following we restrict the study to the period of strong positive correlation between 1999 and 2020. The inclusion of the earlier data leads to inferior results. The possibilities for higher accuracy predictions are best seen on a diagram showing cross-correlation for all months in a year (Fig. 3).

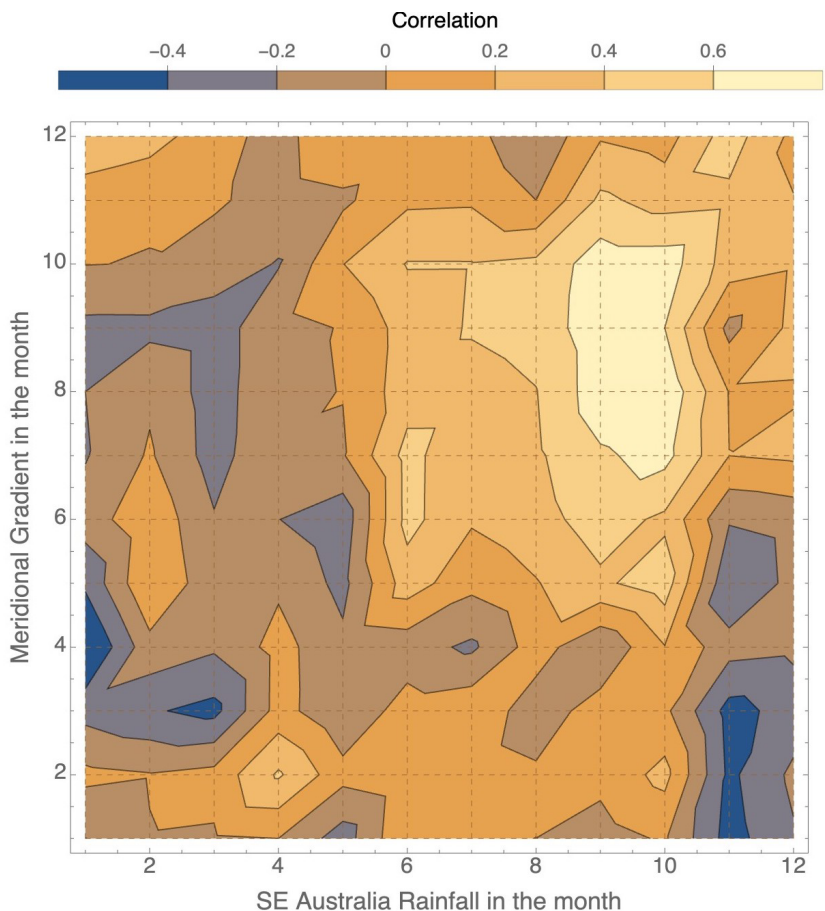

Fig. 3.
Correlation between the Meridional gradient in any month and the rainfall in SE Australia during another month in the same year, evaluated over the period 1999-2020. Note that the high correlation region extends away from the same-month diagonal at the position 9 to 10 towards the months 7 and 8 thus opening a limited window for accurate predictions.

## 3 Neural network hindcasting of rainfall

The data stream input to neural networks was limited to the 1999-2020 period of strong correlations between the rainfall and the Indian Ocean SST. As the favourable period is only 22 years long, the sensitivity of the results to the division of the data into a training set and a validation set is increased. Good compromises, which led to similar success in hindcasting, were obtained by setting the length of the training set to 12-15 years, which left 10 -7 years for the validation set. We used deep learning networks of 3-5 linear layers with a tanh nonlinearity after the first layer. All-linear networks were inferior. The training used a sign of the stochastic gradient or adaptive learning rate (ADAM) optimiser.

An example of the input data stream is shown in Fig. 4. Sometimes it was favourable to use two components of the Meridional gradient separately, or for the case of the South Australian rainfall, vary slightly the extent of the *sI* geographical area.

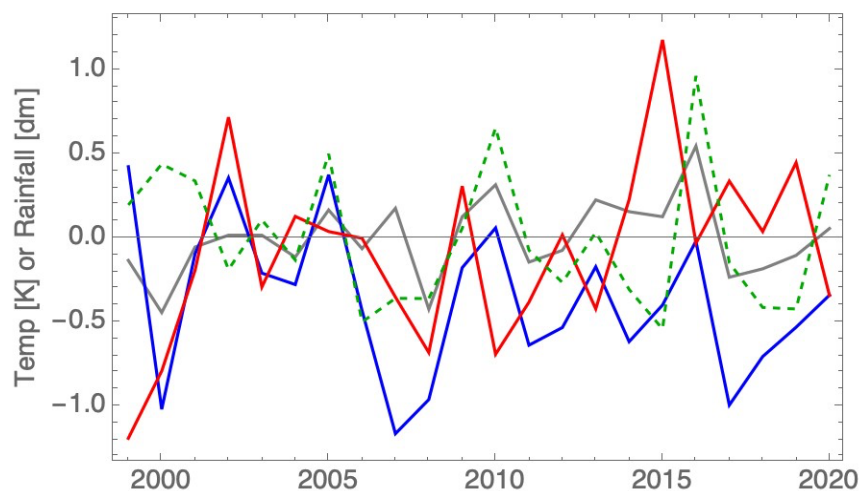

Fig. 4.
Typical neural network input data set optimised for September/October rainfall hindcasting at the end of June. Meridional gradient is shown in blue, East IOD in grey, Niño34 in red and SE Australia rainfall anomaly in a green dashed line. The data used are average monthly values. The time series was divided into the training set until 2010 and the validation set from 2011. At some tasks separate inclusion of the *eI* and *sI* data sets led to more accurate results. Small differences in the division between the training set and the validation set did not lead to significant changes in the hindcast values.

Testing different inputs into the hindcasting networks shows that the springtime rainfall is predominantly determined by the Indian Ocean conditions. The Meridional gradient as introduced Ummenhofer et al. (2009) is more important than the IOD index. We did not use the full IOD index but sometimes used the SST of the eastern input to the IOD. Geographically, best hindcasts are obtained for the average SE Australia rainfall, with Victoria and South Australia somewhat less accurate. In the remainder of this section, we discuss the best use of El Niño information and then present the network results for the validation part of the data.

El Niño cycle is important but plays a secondary role. We illustrate the influence of El Niño in hindcasting spring rainfall with two examples of more difficult hindcasting for South Australia. When hindcasting of September/October rainfall at the end of August, the El Niño time series is less important and, in some cases, may even be omitted with only minor differences in the result (Fig. 5).

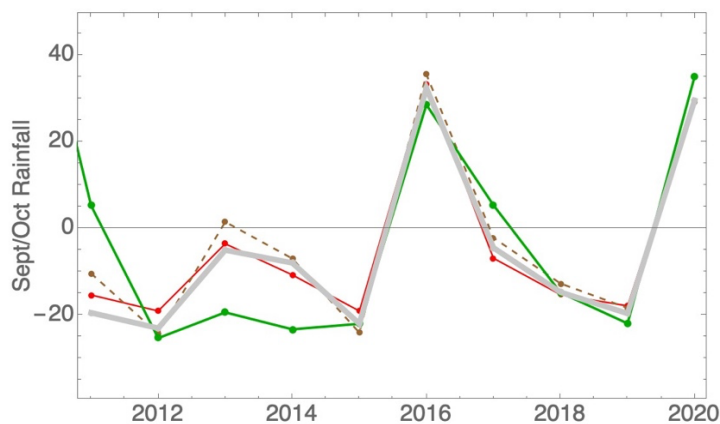

Fig. 5.

Hindcasting average September/October rainfall in South Australia at the end of August. In this and all subsequent figures the rainfall is shown in green, and the period displayed is the validation period showing the performance of the trained neural network.

Two network hindcasts including El Niño data are shown in red and dashed brown. The only input into the third network shown in light grey is the Meridional gradient and SST in the eastern region of the IOD, indicating that Indian Ocean conditions dominate the springtime rainfall.

Predictions of springtime rainfall made two months earlier, from the end of June, are sensitive to El Niño conditions. The relationship of ENSO to weather systems frequency and ultimately the rainfall in SE Australia is very variable with a complicated dynamics (Hauser et al. 2020). In the present neural network study, we only use Niño34 temperatures together with the Indian Ocean inputs, but in a more detailed work the use of a more complete information may shed some light on the underlying physical processes.

Accurate El Niño predictions for the August/September/October period available in June provide an opportunity to improve the neural net performance in hindcasting of the springtime rainfall. The skill of different laboratories in El Niño predictions over the period 2002-11 was evaluated by Barnston et al. (2012) and we selected to use ECMWF historical prediction data available from IRI at Columbia University. An example of the hindcasts using ECMWF predictions for August/September/October period issued in June and compared to those using current June El Niño data is shown in Fig. 6.

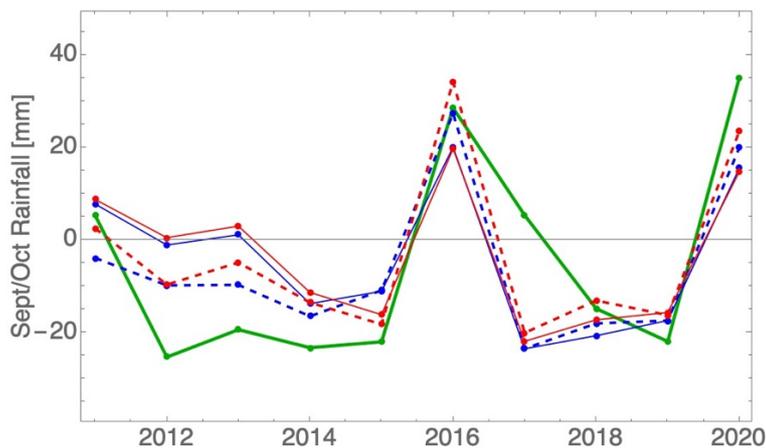

Fig. 6.
An example of the improvements obtained using the forecast value of El Niño anomaly in network inputs.
This example shows a relatively difficult hindcast of the SA rainfall evaluated from the end of June.
The combined September/October rainfall anomaly is shown in green, and hindcasts by two networks are
shown in blue and red. Full lines are obtained with networks trained using the data available at the time of
hindcast. In the same networks, replacing June values for the Niño34 anomaly with the June ECMWF
forecasts for August/September/October Niño34 anomaly led to improved results (dashed lines).

The performance of the best trained networks in hindcasting the validation set data is illustrated
in Figs. 7-8. All rainfall anomalies are defined with respect to the 1999-2020 period, with the
combined September/October rainfall averages of 99.7 mm, 104.4 mm and 32.9 mm for SE
Australia, Victoria and South Australia respectively. Results shown as June etc. are available when
June monthly averages are posted, normally in July. Most of the earlier hindcasts (June and July)
use ECMWF predictions for August/September/October Niño34 area. Hindcasting outside of the
September/October period is significantly worse and becomes useless during the periods of a
year when the correlation is weak (cf. Fig. 3). For hindcasts evaluated in the more interesting
earlier period (using June data) correlations with the actual rainfall are in the range 0.85-0.90,
with the highest values obtained for SE Australia and lowest for SA.

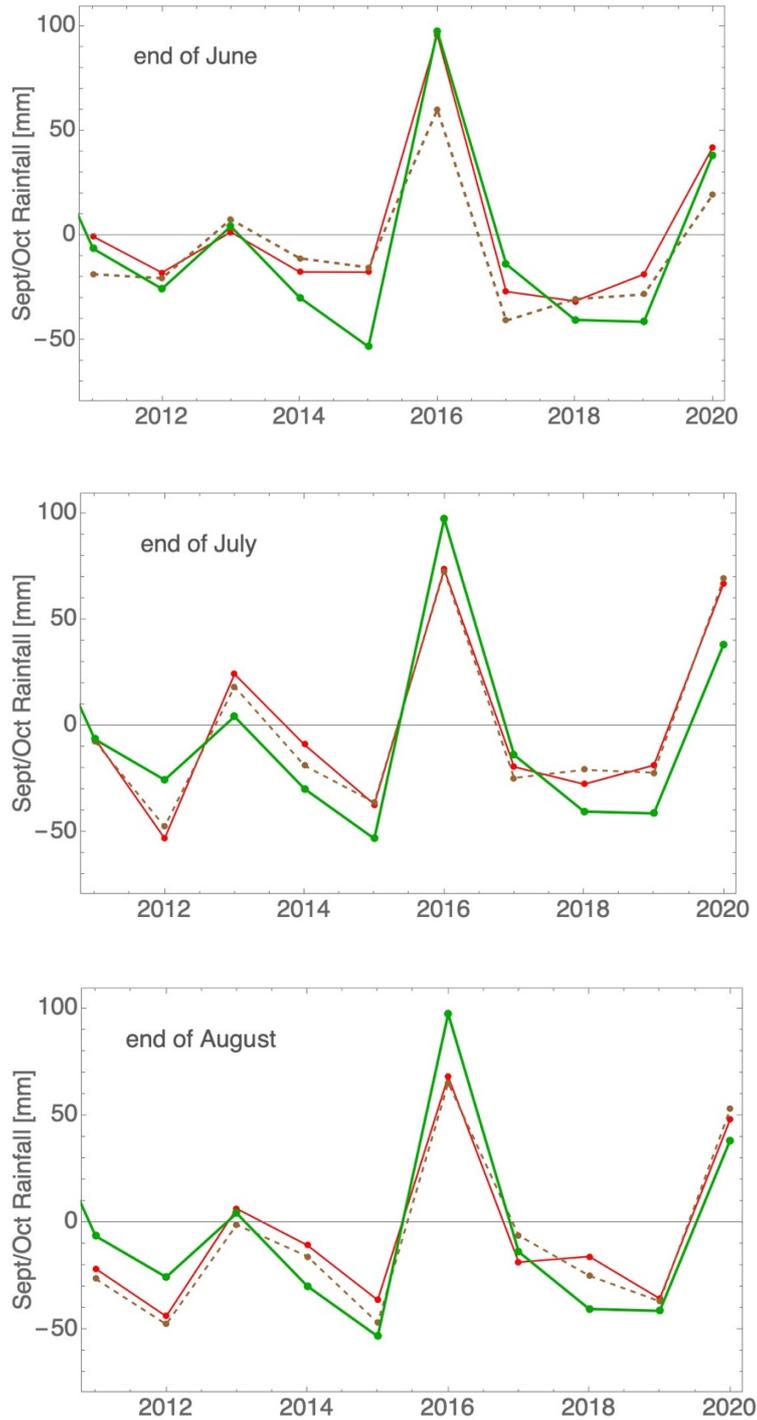

Fig. 7.

Hindcasting results for the cumulative September/October rainfall in SE Australia obtained at the end of each winter month as shown. The period shown is the validation period showing the performance of the trained neural network. The rainfall data are shown in green and two different realisations of neural network hindcasting with slightly different inputs are shown in red and dashed brown.

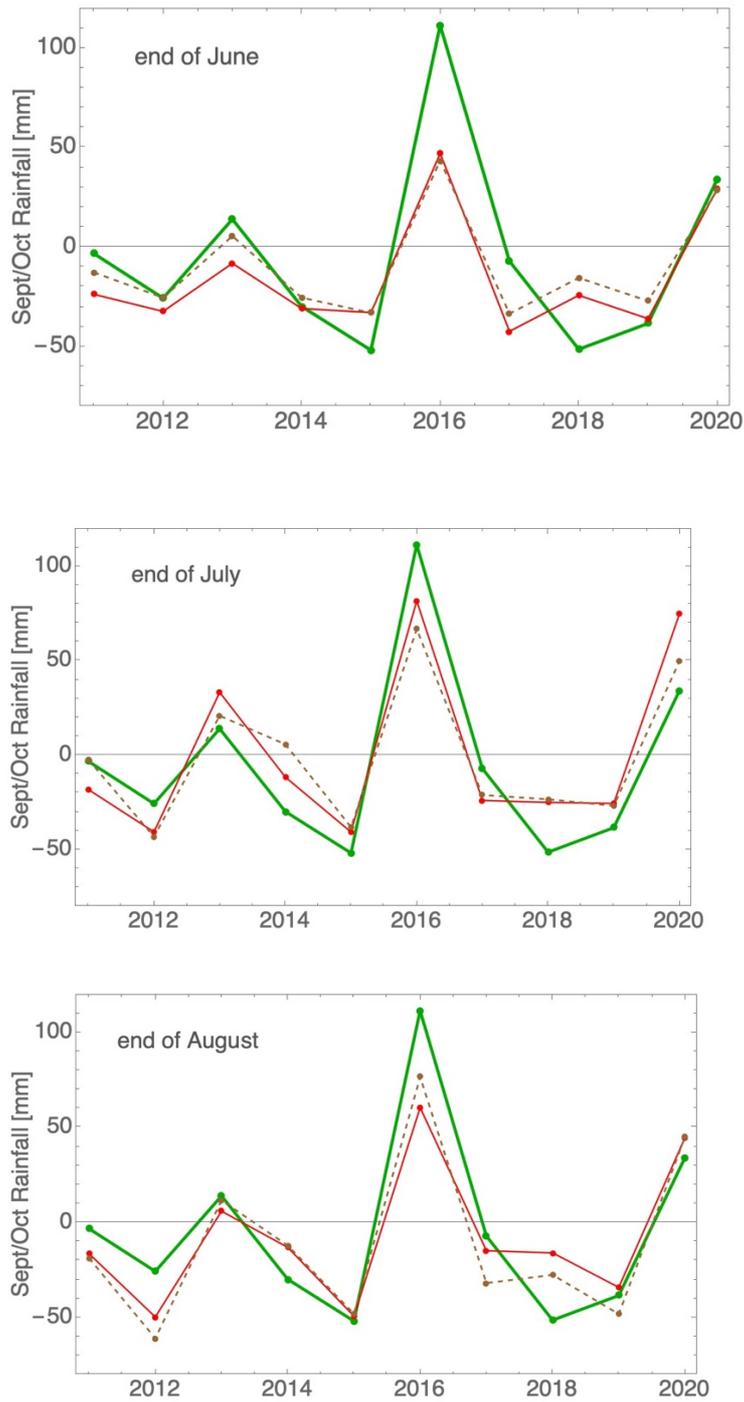

Fig. 8.

Same as Fig. 7 but applied to Victoria. The results are not as accurate, except for hindcasts at the end of August. One of the networks in the August panel (shown as a dashed line) has no input from El Niño, like the example in Fig. 5

**4 Simulated forecasting**

A valid objection to the hindcasting results shown above is that a choice of input data streams, network architecture, training method, learning rate and the moment of terminating the network training is selected to optimise the validation set results. When we move from hindcasting to forecasting, validation values are not available, and the performance will deteriorate. To explore how serious is this bias we made several tests by returning to the years 2018 and 2015. Network training was performed using only the information available in that year. Using networks trained with earlier data, we attempted to forecast springtime rainfall in the future years from the current winter observations.

The networks parameters selected with the diminished validation set were different, but the differences were minor. The forecasting of 2019 and 2020 springtime rain with the networks trained using the data available in 2018 is shown in Fig. 9. Overall performance with the reduced information was not substantially different, indicating that the dominant influence in the results is the information contained in the training set.

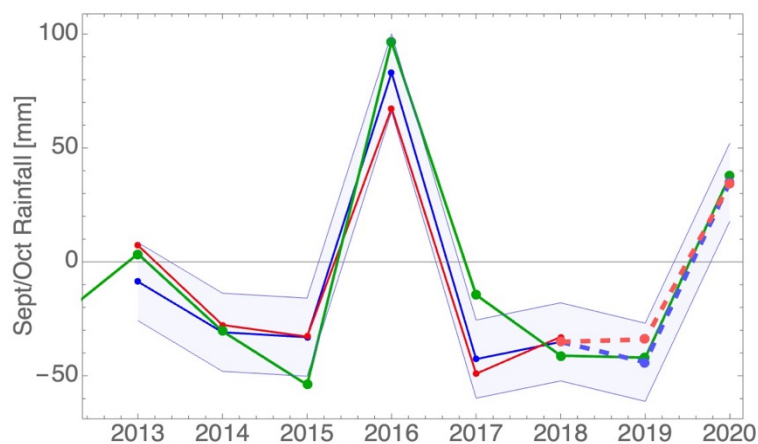

Fig. 9.
Two realisations for the forecast of the spring rainfall anomaly in SE Australia for 2019 and 2020 with the networks optimised using the data available at the end of June 2018. In this test the validation set is shorter, using only 2013-2018 rainfall data.The standard deviation band for one of the network choices is shown with light shading. The mean forecast values for the anomaly were -39±16 mm and +34±16 mm for 2019 and 2020 respectively, while the actual anomaly was -42 mm and +37 mm.

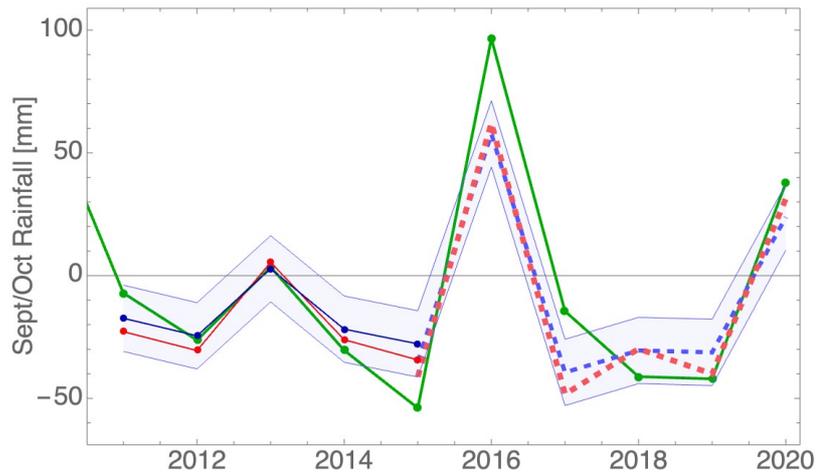

Fig. 10.
Forecasts for the SE Australia spring rainfall anomaly evaluated with the networks trained in June 2015. Two different networks are shown in blue and red, validation set is shown as full lines and forecasts are shown as dashed lines. Of the five forecast years, 2016 and 2017 are outside of the standard deviation limits, which are too small due to a short validation set.

Going further back, we show forecasts using the networks trained with the data available in 2015 (Fig. 10). The already short validation set is further reduced, missing the high rainfall event in 2016. The forecast underestimates the 2016 rainfall, overestimates the drought in 2017 but appears accurate for the next three years. In real applications, one would not use 2015 networks beyond 2016 as the networks improve with every passing year.

Overall accuracy for forecasting in South Australia and Victoria is lower than that for the average over SE Australia. Nevertheless, tests for SA with networks trained in 2018 (Fig. 11) show that despite the decreased accuracy, in a combined hindcast/forecast result the general trend of predicting the rainfall higher or lower than the average is missed only once in 10 years.

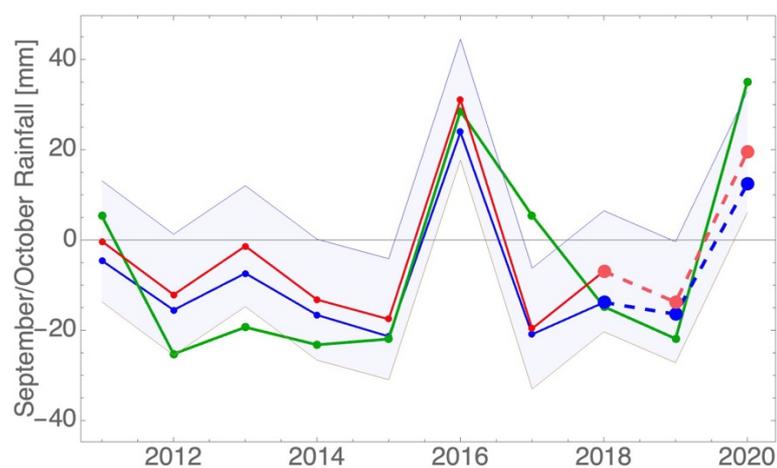

Fig. 11.

Testing the forecasting skill for average springtime rain in the more difficult case of South Australia with the networks trained at the end of June 2018. The result is similar to the hindcasts shown in Fig. 6.

**5 Discussion**

The present study confirmed the dominant influence of strong teleconnections with the Indian Ocean SST on the rainfall in SE Australia. The teleconnections strengthened within the last two decades as seen in the time dependence of the correlations (Fig. 2). The change might be driven by the Pacific Decadal Oscillation (Lim et al. 2016a), which could mean that it can reverse in the future. Alternatively, the climate change which leads to slow poleward drift of climate zones may be responsible for the permanently stronger link of SE Australian climate with the state of tropical Indian Ocean.

A strong feature is visible in the cross-correlation plot of Fig. 3. Specific regions of the Indian Ocean during the winter months contain the information about the future springtime rainfall, but such link appears weak or non-existent at other times of the year. The findings of this statistical analysis point to seasonal physical processes that could be identified and incorporated into the more comprehensive modelling.

We used deep learning neural network method to search ocean surface temperatures for the information relevant to future climate events. While large dynamical models like ECMWF or ACCESS ultimately perform better than statistical models (Barnston *et al*. 2012, for the case of ENSO) neural network approach proved useful in identifying the features that need to be present and understood within the dynamical models. Networks may be useful in exploring untapped predictability that was not recognised in dynamical models. For example, it is known that ACCESS-S1 has systemic weakness in representing teleconnection between climate drivers and rainfall in eastern and SE Australia, particularly during the winter season (Hudson et al. 2017). In the present study, neural network approach identified Indian Ocean meridional gradient as a stronger predictor of springtime rain in SE Australia than the generally recognised climate drivers IOD and ENSO. More generally, neural network methods may provide help in improving dynamical models to fully exploit the predictability potentially available in different climate data.

A novelty in the approach of this study is a strictly practical combination of using the results from one of the best dynamical models for El Niño forecasting (ECMWF) together with the current SST data as an input into the neural networks. This hybrid strategy of simultaneously using information from both dynamical and statistical modelling may prove helpful in overcoming specific obstacles in forecasting tasks.

The neural networks used here are examples of unsupervised learning, where no information is provided on the physical significance or the importance of each input data stream. Such methods sometimes prove valuable in solving complex tasks that involve large number of possible states. Presently, we do not know to what extent the practical prediction of springtime rain in SE Australia will ultimately be improved with the aid of artificial intelligence. At the time of writing, late July 2021, the average SST values for the month of June are available. Using the simple network examples obtained with June 2020 data and shown in Figs. 5-8 we expect that the total rainfall during the September/October 2021 period will be close to the average. The anomaly is forecast as -6±15 mm in SE Australia, -5±27 mm in Victoria and -6±12 mm in South Australia.

*Addendum 29 July 2022*

The neural network forecasting is improved by taking the average over many training sessions that start with the rate of network weights adjustments randomly chosen between the selected limits. The rate must be large enough to enable the escape from shallow local minima and small enough to enable the convergence of a network to a stable configuration.

The results are illustrated in Fig A1 by forecasting the rainfall anomaly for SE Australia during the next September/October period.

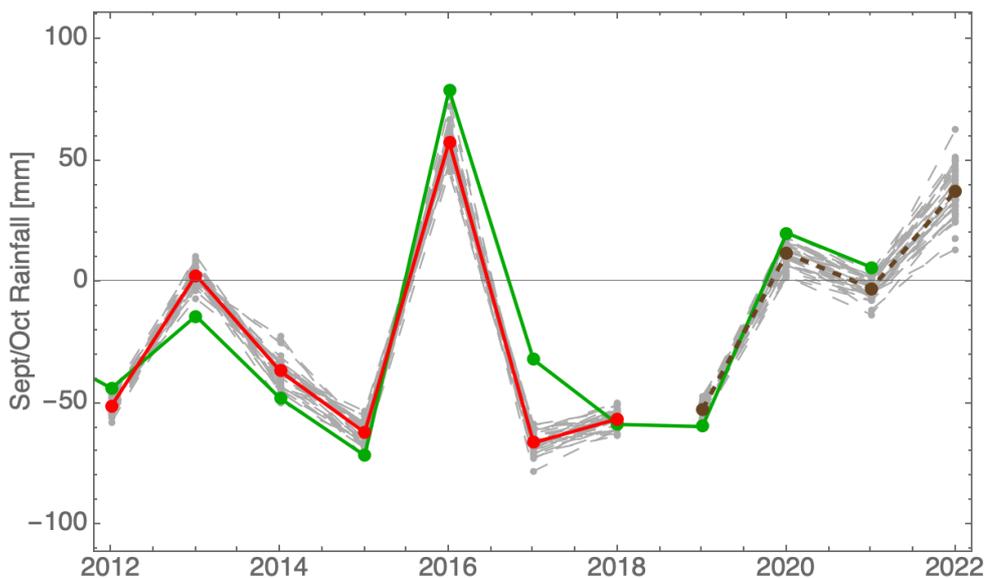

Fig A1

Forecasting SE Australia rainfall anomaly for the September/October period at the end of July. Past rainfall is shown in green, the mean values for the validation period in red and the hindcast (2019- 2021) and forecast (2022) in black. Hindcasting springtime rainfall is the only time when accurate results were obtained.

The forecasts obtained for the 2022 September/October period rainfall anomaly are:

Southeastern Australia:    38 ± 8 mm (Fig A1)
Victoria:                          46 ± 12 mm

The reference values for the average rainfall are those used by the Australian Bureau of Meteorology (1961-1990, 118.2 and 129.1 mm for SE Australia and Victoria, respectively).

**Declaration of Funding**
This research did not receive any specific funding.
**Acknowledgements**
All data sources as well as the use of NOAA PyFerret program ([http://ferret.pmel.noaa.gov/Ferret/](http://ferret.pmel.noaa.gov/Ferret/)) are gratefully acknowledged. Very valuable suggestions from Vjeko Matić and Barry Ninham are highly appreciated.

**REFERENCES**

Barnston, A. G, Tippett, M. K, L'Heureux, M., L, Li, S. and deWitt, D.G. (2012). Skill of real-time seasonal ENSO model predictions during 2002-11. Bull. Am. Met. Soc. **93**, 631-651. doi:10.1175/BAMS-D-11-00111.1

Cai, W. and Cowan, T. (2008). Dynamics of late autumn rainfall reduction over southeastern Australia. *Geophys. Res. Lett.* L09708. doi:10.1029/2008GL033727

Cai, W., van Rensch, P., Cowan, T. and Hendon, H. H. (2011a). Tele-connection pathways of ENSO and the IOD and the mechanisms for impacts on Australian rainfall. J. Clim. **24**, 3910–3923. doi: 10.1175/2011JCLI4129.1

Cai, W., van Rensch, P., Cowan, T. (2011b). Influence of Global-Scale Variability on the Subtropical Ridge over Southeast Australia. J. Clim. **24**, 6035–6053. doi: 10.1175/2011JCLI4149.1

Han, W., Vialard, J., J. McPhaden, M. J., Lee, T., Masumoto, Y., Feng, M. and de Ruijter, W. P. M (2014). Indian ocean decadal variability: A Review. Bull. Am. Meteor. Soc. **95**, 1680-1703. doi:10.1175/BAMS-D-13-00028.1

Hauser, S., Grams, C.M., Reeder, M.J., McGregor, S., Fink, A.H., Quinting, J.F. (2020). A weather system perspective on winter–spring rainfall variability in southeastern Australia during El Niño. *QJR Meteorol. Soc*. **146**, 2614– 2633. doi: 10.1002/qj.3808

Hudson, D., Alves, O., Hendon, H., Lim, E.-P., Liu, G., Luo, J.-J., MacLaughlan, C., Marshall, A. G., Shi, L., Wang, G., Wedd, R., Young, G., Zhao, M., and Zhou, X. (2017). ACCESS-S1: The new Bureau of Meteorology multi-week to seasonal prediction system. *J. South. Hemisph. Earth. Sys. Sci.* **67**, 132–159. doi: 10.22499/3.6703.001

Lim, E.-P, Hendon, H.H., Zhao, M. and Yin, Y. (2016a). Inter-decadal variations in the linkages between ENSO, the IOD and south-eastern Australian springtime rainfall in the past 30 years**.** *Clim. Dyn.,* **49**, 97–112. doi: 10.1007/s00382-016-3328-8

Lim, E.-P, Hendon, H.H., Hudson, D., Zhao, M., Shi, L., Alves, O. and Young G. (2016b). Evaluation of the ACCESS-S1 hindcasts for prediction of Victorian seasonal rainfall. Bureau Research Report No. 19, Bureau of Meteorology Australia (available from: http://www.bom.gov.au/research/research-reports.shtml).

Nooteboom, P. D., Feng, Q. Y. López, C., Hernández-García, E, and Dijkstra, H. A. (2018). Using network theory and machine learning to predict El Niño. *Earth Syst. Dynam.*, 9, 969–983. doi: 10.5194/esd-9-969

Ratnam, J. V., Dijkstra, H. A. and Behera, S. K. (2020). A machine learning based prediction system for the Indian Ocean Dipole. *Sci. Reports* **10**, 284. doi: 10.1038/s41598-019-57162-8

Timbal, B. and Drosdowsky, W. (2012). The relationship between the decline of Southeastern Australian rainfall and the strengthening of the subtropical ridge. *Int. J. Climatol.* **33**, 1021-1034. doi: 10.1002/joc.3492

Ummenhofer, C. C., Sen Gupta, A., Pook, M. J. and England, M. H. (2008). Anomalous Rainfall over Southwest Western Australia Forced by Indian Ocean Sea Surface Temperatures. *J. Clim.* **21**, 5113-5134. doi: 10.1175/2008JCLI2227.1

Ummenhofer, C. C., Sen Gupta, A., Taschetto, A. S. and England, M. H. (2009). Modulation of Australian Precipitation by Meridional Gradients in East Indian Ocean Sea Surface Temperature. *J. Clim.* **22**, 5597-5610. doi: 10.1175/2009JCLI3021.